\documentclass[journal]{IEEEtran}
\usepackage{amsmath,amsfonts}
\usepackage{algorithmic}
\usepackage{algorithm}
\usepackage{array}
\usepackage[caption=false,font=normalsize,labelfont=sf,textfont=sf]{subfig}
\usepackage{textcomp}
\usepackage{multirow}
\usepackage{stfloats}
\usepackage{url}
\usepackage{bbding}
\usepackage{booktabs}
\usepackage{verbatim}
\usepackage{amsmath} 
\usepackage{amssymb}
\usepackage{graphicx}
\usepackage{cite}
\usepackage{color}
\begin{document}

\title{BAZAM: A Blockchain-Assisted Zero-Trust\\ Authentication in Multi-UAV Wireless Networks}

\author{Mingyue Xie, Zheng Chang,~\IEEEmembership{Senior~Member,~IEEE}, Osama Alfarraj, Keping Yu,~\IEEEmembership{Senior~Member,~IEEE},\\Tao Chen, Hongwei Li,~\IEEEmembership{Fellow,~IEEE}
	   \thanks{M. Xie, Z. Chang, T. Chen, and H. Li are with School of Computer Science and Engineering, University of Electronic Science and Technology of China, Chengdu 611731, China (e-mail: mingyueyue@std.uestc.edu.cn, zheng.chang@jyu.fi, chent@uestc.edu.cn, lihongwei@uestc.edu.cn). }
    
    \thanks{Osama Alfarraj is with Computer Science Department, Community College, King Saud University, Riyadh 11437, Saudi Arabia (e-mail: oalfarraj@ksu.edu.sa). }
    
    \thanks{Keping Yu is with Graduate School of Science and Engineering, Hosei University, Tokyo 184-8584, Japan (e-mail: keping.yu@ieee.org).}
 
 
 }
\markboth{Journal of \LaTeX\ Class Files}%
{Shell \MakeLowercase{\textit{et al.}}: A Sample Article Using IEEEtran.cls for IEEE Journals}

 
\maketitle

\begin{abstract}
Unmanned aerial vehicles (UAVs) are vulnerable to interception and attacks when operated remotely without a unified and efficient identity authentication. Meanwhile, the openness of wireless communication environments potentially leads to data leakage and system paralysis. However, conventional authentication schemes in the UAV network are system-centric, failing to adapt to the diversity of UAVs identities and access, resulting in changes in network environments and connection statuses. Additionally, UAVs are not subjected to periodic identity compliance checks once authenticated, leading to difficulties in controlling access anomalies. Therefore, in this work, we consider a zero-trust framework for UAV network authentication, aiming to achieve UAVs identity authentication through the principle of ``never trust and always verify''. We introduce a blockchain-assisted zero-trust authentication scheme, namely BAZAM, designed for multi-UAV wireless networks. In this scheme, UAVs follow a key generation approach using physical unclonable functions (PUFs), and cryptographic technique helps verify registration and access requests of UAVs. The blockchain is applied to store UAVs authentication information in immutable storage. Through thorough security analysis and extensive evaluation, we demonstrate the effectiveness and efficiency of the proposed BAZAM.

\end{abstract}


\section{Introduction}
\subsection{Background}
In the next-generation networking paradigm, unmanned aerial vehicles (UAVs)-based wireless networks, benefiting from the rapid and easily deployable characteristics\cite{int1}, have been extensively applied across various domains \cite{int2}. Due to the deployment in open environments and reliance on wireless communication channels, UAVs are susceptible to interception and attacks\cite{int3}, such as man-in-the-middle attacks, replay attacks, impersonation attacks, and physical capture attacks\cite{int4, int5}. These attacks have the potential to disrupt the normal operation of UAVs, allowing attackers to illicitly obtain resources from internal networks. Therefore, secure UAV authentication technologies play a pivotal role in effectively curbing adversarial attacks\cite{int6}.

However, due to the characteristics of multi-UAV wireless networks, it is challenging to apply the conventional network security model and scheme directly to design authentication schemes of UAV networks\cite{int7}. On one hand, the high-speed mobility of UAVs leads to rapid changes in network conditions and connection status, necessitating the repeated verification of their identities\cite{int8, int9}. On the other hand, an excessive trust in UAVs within the internal network makes the system difficult to effectively defend against internal malicious UAVs\cite{int10}. Moreover, traditional authentication approaches primarily relying on tag, passwords, tokens, and fingerprints are susceptible to theft or forgery by attackers\cite{int11}, thereby enabling unauthorized access to internal network resources. Additionally, these approaches typically grant prolonged access privileges to UAVs after authentication, leading internal network resources leakage. While boundary-based multi-UAV wireless network protection can effectively thwart external attacks, it cannot prevent attacks initiated by malicious UAVs within the internal network. Consequently, conventional multi-UAV wireless network architectures still face significant security challenges.

A novel network security architecture known as zero-trust\cite{int12}, shows its potential to be an appropriate solution for UAV authentication. The zero-trust model posits that both the internal and external aspects of wireless networks are untrusted\cite{int13, int14}. In other words, in the zero-trust network, UAVs are deemed untrusted until thoroughly validated and authenticated. Furthermore, the zero-trust model aims to transform traditional network architectures from being system-centric to user-centric with continuous authentication \cite{int15}. When a UAV needs wireless or other resources, its identity is initially authenticated and assessed to determine its access permissions\cite{int16}. Subsequently, the zero-trust gateway allows or denies UAVs access to service resources and monitors their access behaviour in real time. The zero-trust network protection shifts from the network perimeter to the user identity itself, effectively preventing unauthorized access requests from within the network. For multi-UAV wireless networks, the zero-trust network architecture can provide a higher level of security protection, safeguarding sensitive data from unauthorized access and potential security threats.

\subsection{Related Work}
Recently, considerable efforts have been dedicated to the research of security and threat towards UAVs. In UAV networks, the changes in topology may occur more frequently \cite{rel1}. The relative positions of UAVs may undergo alterations, leading to the formation and disappearance of communication links. Additionally, UAVs may experience malfunctions, resulting in disconnection from the network \cite{rel2}. Tiburski et al. \cite{rel3} mention that if attackers gain physical access to a UAV and tamper with its authentication parameters, the legitimate UAV in the network may be mistakenly identified as malicious. Wang et al.\cite{adrel1}  claim that attackers may attack open wireless channels, potentially engaging in illicit interception of communication messages, and even manipulating the dissemination of messages through forgery and control. In \cite{rel4}, attackers may disrupt, intercept, or interfere with the communication signals of UAVs, thereby impacting the control of UAVs or data transmission. Once attackers eavesdrop on the private communication of a UAV and then sniff its sensitive information, such as identity, private key, passwords, etc., they can impersonate the UAV, leading to unauthorized access to the network\cite{rel5,rel6}.

Several cryptographic frameworks have been proposed for UAV authentication in recent years. In \cite{rel7}, elliptic curve cryptography (ECC) is utilized to design a secure authentication scheme for UAV communications, including message authentication and key agreement. UAVs can authenticate independently instead of relying on a key generation center. In \cite{rel8}, a distributed UAV authentication scheme is designed based on blockchain. The authentication process is based on the ECC, and the UAVs can invoke the smart contract to update its information in the blockchain to achieve authentication. Considering the parameter sizes of ECC, the authors in \cite{rel9} propose a provable and privacy-preserving UAV authentication scheme based on hyperelliptic curve cryptography, reducing the cost caused by processing parameters in UAV authentication. In \cite{rel10}, a lightweight mutual authentication protocol for UAVs is designed. It is solely comprised of bitwise operations and one-way hash functions. Another popular approach in UAV authentication is to leverage physical layer attributes. In \cite{rel11}, a lightweight authentication scheme is proposed based on the physically unclonable function (PUF). It supports UAV-ground station (GS) and UAV-UAV communication modes and is resistant to physical capture and node tampering attacks. To ensure physical security and fast authentication, the authors in \cite{rel12} also apply the PUF to design a mutual authentication protocol for UAV swarms. In \cite{rel13}, the PUF and chaotic system to achieve mutual authentication are employed to establish a secure session key between communication UAVs. In \cite{rel14}, a PUF and fuzzy extractor-based mutual authentication mechanism is designed for UAV-GS. Both UAVs and the GS are unable to store any secret information that can be used to decrypt messages communicated between them.

There have been increasing interests in authentication using zero-trust. In \cite{rel15}, a security awareness and protection system that leverages zero-trust architecture is proposed, which achieves continuous identity authentication, analysis of access behavior, and fine-grained access control. The authors in \cite{rel16} develop a zero-trust software-based security framework to provide secure communications within the mobile core network by using an authentication-based approach. In the zero-trust environment, only authenticated and authorized entities can access each other. In \cite{rel17}, a blockchain-enabled information sharing solution is designed in zero-trust context to guarantee entity authentication. Applying the smart contracts and consensus mechanisms, unauthenticated participants are prevented from sharing information. The authors in \cite{rel18} realize the ownership authentication and operation authorization of UAVs through the incorporation of zero-trust architecture into the blockchain and the artificial intelligence models. The authors in \cite{rel19} construct a zero-trust authentication framework based on Markov games with one-sided information, which leverages interdependent trust evaluation to automate zero-trust security for the fifth generation Internet of Things (IoT) networks. In \cite{rel20}, a three-step physical layer enhance zero-trust security framework for wireless industrial IoT, the devices authentication of which is based on the physical fingerprint, eliminating third-party trusted certification authorities. In \cite{rel21}, the authors propose a holistic research agenda for zero-trust user authentication in social virtual reality, which investigates biometrics-based authentication for continuously authenticating users and improves the authentication accuracy with multimodal data.

\subsection{Motivation and Contributions}
As mentioned above, existing UAV wireless networks suffer from potential security attacks. In this context, identity-based continuous authentication is important to improve the multi-UAV network security. Taking into account the dynamics of UAV network, as well as the intranet trust and security of authentication approaches, the zero-trust model can establish a robust and adaptable UAV authentication scheme. Therefore, this paper proposes a blockchain-assisted authentication scheme, called BAZAM, for multi-UAV wireless networks under the zero-trust framework. The main contributions of this paper are summarized as follows:

\begin{itemize}
\item We propose a blockchain-assisted UAV authentication scheme specifically designed to operate in the zero-trust network. The zero-trust network centers around the UAV identity, which is authenticated and evaluated by the software defined perimeter (SDP) controller before allowing resource access requests from a UAV. 
\item We utilize the randomness and non-replicability on the UAV hardware to generate its secret key. Based on the PUF, we present a lightweight cryptographic technique as identity authentication to design the BAZAM. The SDP controller can verify the identity and request of a UAV before sending a policy to the gateway to allow UAV access. 
\item To enhance the authentication efficiency of the SDP controller in the zero-trust framework, we introduce a reputation to filter UAVs' eligibility for authentication preparation, taking into account the legitimate authentication attempts made by the UAVs. 
\item We introduce a blockchain-assisted authentication service for UAVs. The reputation required for the authentication is stored in the blockchain. The untamperable record increases the reliability and security of reputation and reduces the risk of single points of failure and centralization in zero-trust network. 
\end{itemize}

The remainder of this paper is organized as follows. Section II presents the basic concepts. We describe the system model, threat model, security goals and adversary model in Section III. The detailed constructions of the proposed scheme are shown in Section IV. Section V and VI demonstrates the security analysis and performance analysis. Section VII concludes the research work.

\section{Preliminaries}
In this section, we introduce the following concepts for the construction of our scheme. 
 
\subsection{Intractable Problem}


As the intractability assumptions of the hard problems is the theoretical basis for the later proof, we utilize a slight modification presented in \cite{adpre1} to provide a security basic for our BAZAM scheme.

\textit{Definition 1 (Modified Bilinear Diffie-Hellman (MBDH) assumption)}: Given a cyclic additive group $\mathbb{G}_1$ and a cyclic multiplicative group $\mathbb{G}_2$ with the same order $q$, a bilinear map $e:\mathbb{G}_1 \times \mathbb{G}_1 \xrightarrow{} \mathbb{G}_2$, and a generator $G$ of $\mathbb{G}_1$, for a tuple ($G$,$aG$,$bG$,$cG$,$c^{-1}G$), $z$$\in$$\mathbb{G}_2$, $a$, $b$, $c$ are unknown integers, MBDH assumption is to compute the value of $e(G, G)^{abc}$.

\textit{Definition 2 (Modified Decisional Bilinear Diffie-Hellman (MDBDH) assumption)}: Given a cyclic additive group $\mathbb{G}_1$ and a cyclic multiplicative group $\mathbb{G}_2$ with the same order $q$, a bilinear map $e:\mathbb{G}_1 \times \mathbb{G}_1 \xrightarrow{} \mathbb{G}_2$, and a generator $G$ of $\mathbb{G}_1$, for a tuple ($G$,$aG$,$bG$,$cG$,$c^{-1}G$), $z$$\in$$\mathbb{G}_2$, $a$, $b$, $c$ are unknown integers, MDBDH assumption is to decide whether $z$ = $e(G, G)^{abc}$.

\subsection{Physical Unclonable Function}
The PUF refers to a physical entity that responds with inputs and outputs \cite{pre1}. This response depends on randomly varying physical structures, as well as mismatches resulting from the intrinsic structure of the physical entity. Compared to conventional encryption algorithms, the implementation of PUF is relatively straightforward and offers higher level of security. In PUF, challenge-response pairs constitute a pivotal concept employed for device authentication or key generation. A PUF can be defined as the function $r_n$ = PUF($c_n$). A set of $n$ input values (referred to as the challenge) $C$ $\in$ $\{$$c_0$, ..., $c_n$$\}$ representing a randomly or specifically input, typically manifests as electrical signals or data. Subsequent to challenge generation, it is conveyed to the PUF system. Upon receipt of a challenge, the PUF system, influenced by its internal physical structure and inherent randomness, engenders a singular response $R$ $\in$ $\{$$r_0$, ..., $r_n$$\}$. This response is a distinct reaction to a challenge, thereby endowing each PUF system with a unique response pattern.  

\subsection{Software Defined Perimeter}
The conventional network security models typically rely on static boundary defenses, such as firewalls and virtual private networks, which lack adequate flexibility and security. Therefore, the technological system in the zero-trust architecture is software defined perimeter (SDP) \cite{pre2}. The typical SDP architecture consists of SDP controller, SDP client, and SDP gateway. The SDP controller is responsible for validating packet sent by the SDP client, issuing policies, managing the SDP gateway, and configuring the connection between the SDP client and SDP gateway. The SDP client, a software or agent on user devices, is tasked with communicating with the SDP controller for identity verification and access control. The SDP gateway, located at the network edge, by default denies all access requests. Only upon successful authentication of a request and the issuance of corresponding credentials and policies by the SDP controller, the SDP gateway establishes a connection with the SDP client.



\section{System Model and Problem Formulation}

In this section, we present the system model of BAZAM, and formulate the threat model, security goals, and adversary model.

\subsection{System Model}

\begin{figure}[!t]
\centering
\includegraphics[width=\linewidth]{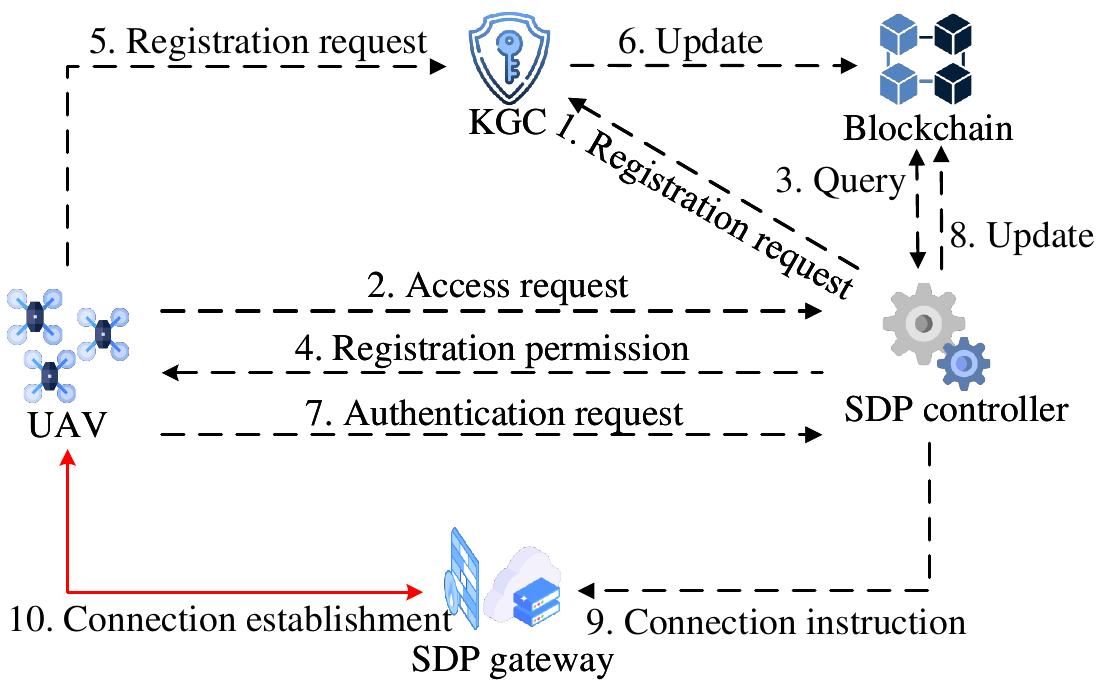}
\caption{System model of BAZAM.}
\label{fig:Architecture}
\end{figure}

In this paper, we propose BAZAM, a zero-trust multi-UAV authentication scheme based on blockchain, which combines blockchain, cryptography and SDP technology to authenticate a UAV access request, so as to achieve fine-grained authentication and authorisation of UAVs and ensure the security of service resources. To demonstrate the applicability of BAZAM in zero-trust architecture, we suggest the proposed infrastructure that can be divided into five types of essential units with distinctive functionalities: the key generation center (KGC), the UAV, the SDP controller, the SDP gateway, and the blockchain, as shown in Fig. \ref{fig:Architecture}. 

\begin{itemize}
\item{\textit{KGC}: KGC is a trusted party authority. It possesses robust communication and computational capabilities. Additionally, it is responsible for facilitating the registration of UAVs and SDP controller in the zero-trust model, as well as assisting in the authentication process between them. The registration processes of UAVs and SDP controller with KGC occur over a secure channel. Moreover, KGC is deemed entirely trustworthy and impenetrable.}
\item{\textit{UAV}: UAV is assume the role of data resource requester. Each UAV is equipped with a PUF. Upon submitting a resource access request to the SDP controller and receiving registration approval, the UAV proceeds to request its public/ private key pair from the KGC. Subsequently, the UAV utilizes its own private key and the public key of the SDP controller to encrypt and sign the single page application (SPA) packet. Then it presents an authentication request to the SDP controller with the ciphertext. Upon successful authentication, the UAV can establish a secure connection with the SDP gateway, thereby gaining access to the resources.}
\item{\textit{SDP Controller}: SDP controller is utilized to manage and control access to network resources. The primary role of the SDP controller is to dynamically determine, based on the UAV identity and packet information, that only authenticated and authorized UAV has permission to access resources. Additionally, the SDP controller considers the reputation and the corresponding thresholds of the UAV to decide whether to grant registration rights to the UAV.}
\item{\textit{SDP Gateway}: When receiving a request from a UAV, the SDP gateway determines, based on the authorization policy and authentication result provided by the SDP controller, whether to allow the request. If the request is granted, the SDP gateway establishes a secure connection, enabling the UAV to securely access the required resources.}
\item{\textit{Blockchain}: Blockchain is a shared distributed ledger, that consists of immutable and trustworthy information of UAVs. Blockchain is critical for SDP controller to process the registration requests and enforce UAV authentication.}
\end{itemize}

In Fig. \ref{fig:Architecture}, BAZAM works with the following steps for authenticating the UAVs and establishing the connection between the UAVs and the gateway. Firstly, the SDP controller initiates a registration request to the KGC to obtain its own key pairs. Subsequently, the UAV sends an access request to the SDP controller. The SDP controller determines the eligibility of the UAV registration by querying the UAV information on the blockchain. Upon receiving registration permission, the UAV sends a registration request to the KGC to generate a unique public/private key pair, while the KGC updates the UAV's information on the blockchain. Following this, the UAV generates packet data and sends an authentication request to the SDP controller. After verifying the packet and confirming the identity of the UAV, the SDP controller updates the UAV information on the blockchain for subsequent authentication. Upon successful authentication, the SDP gateway receives connection instruction from the SDP controller, allowing it to open corresponding ports for the UAV and grant access to the relevant service resources.

\subsection{Threat Model}
The formalization of the adversary model is based on Dolev–Yao threat framework\cite{thr1}. Based on the security threats faced by UAVs in the process of data transmission, the adversary in the process of identity authentication of UAVs is defined. The adversary has the following capabilities:

When the sensitive data is exchanged over the UAV network, we assume that an adversary might employ sniffing tools to eavesdrop on the packets, thereby gaining access to the communication content. Additionally, an adversary could intercept messages sent from UAVs, forging different messages to be sent to the SDP controller, executing man-in-the-middle attack. Furthermore, an adversary may illicitly obtain secret information stored within UAVs, involving unauthorized access to confidential data stored internally. In the zero-trust network, an adversary has the capability to send messages to all entities, allowing them to impersonate UAVs within the network and send deceptive messages. More critically, an adversary even has the potential to acquire the private key of a UAV. However, an adversary lacks the ability to control the trusted KGC, and it cannot obtain the system's random numbers through guessing.

\subsection{Security Goals}
To fulfil the resource access requirements of multi-UAV in wide-area airspace, and to establish secure communication links between UAVs and zero-trust networks, the access authentication scheme for UAVs within the zero-trust architecture should achieve the following security goals:

\textbf{Mutual Authentication}: BAZAM should ensure that only legitimate UAVs can complete authentication with the SDP controller and defend against adversaries posing as legitimate UAVs to launch malicious attacks.

\textbf{Session Key Establishment}: BAZAM should generate unique session keys for the authentication process. In addition, it should ensure that an adversary is unable to obtain intelligence from the captured session key.

\textbf{Integrity}: BAZAM should perform the verification on the packets origin, and ensure that the data are free from deliberate or inadvertent unauthorized manipulation or modification.

\textbf{Forward and Backward Secrecy}: Even if adversaries obtain the master key or secrets, they should not be able to learn previous session keys established between the UAV and the SDP controller. Additionally, the newly joined UAVs are prevented to learning previous session keys.

\textbf{Secure Against Cyber Attacks}: BAZAM should be secure against common security attacks, such as message tampering, replay attack, etc.

\textbf{Secure Against Physical Attacks}: The adversaries may attempt to create replicas of legitimate UAVs through cloning attacks, as well as physically tamper with UAV identifiers or data in memory. Therefore, BAZAM should be secure against UAV cloning attacks and other physical attacks, such as UAV capture and tampering.

\textbf{UAV Traceability}: The authentication history of any registered UAV can be traced. In other words, the authentication information and results of UAV need to be accurately recorded.

\subsection{Adversary Model}
As proof that our BAZAM scheme satisfies confidentiality and the unforgeability, we provide the security games for our BAZAM scheme.

\textit{Definition 3 (Indistinguishability Against Adaptive Chosen-Ciphertext Attack (IND-CCA2))}: Our scheme is IND-CCA2 secure if no adversary $\mathcal{A}$ possesses a non-negligible advantage in winning the following game.

\textbf{Initial}: A challenger $\mathcal{C}$ initializes the system parameters and master key. Then the parameters are forwarded to $\mathcal{A}$.

\textbf{Phase 1 and 2}: $\mathcal{A}$ adaptively asks to $\mathcal{C}$ the following the queries.
\begin{itemize}
\item{Key extract queries}: $\mathcal{A}$ submits key extract queries for identities of the UAV and SDP controller. $\mathcal{C}$ sends the corresponding private keys to $\mathcal{A}$.
\item{Signcrypt queries}: $\mathcal{A}$ chooses the identities of the UAV and SDP controller and a packet. $\mathcal{C}$ obtains the private key of the UAV for signcrypting the packet and sends it to $\mathcal{A}$.
\item{Unsigncrypt queries}: $\mathcal{A}$ chooses the identities of the UAV and SDP controller and a ciphertext. $\mathcal{C}$ obtains the private key of the SDP controller for decrypting the ciphertext and sends it to $\mathcal{A}$.
\end{itemize}

\textbf{Challenge}: $\mathcal{A}$ adaptively asks to $\mathcal{C}$ the aforementioned queries. When the Phase 1 is finished, $\mathcal{A}$ generates two equal-size messages and the challenging identities of the UAV and SDP controller to $\mathcal{C}$. $\mathcal{C}$ randomly selects $\gamma$ $\in$ $\{ 0, 1\}$ to make the challenging ciphertext and forwards it to $\mathcal{A}$.

\textbf{Guess}: $\mathcal{A}$ outputs a guess $\gamma'$ and wins this game if $\gamma'$ = $\gamma$.

\textit{Definition 4 (Existential Unforgreability Against Chosen-Message Attack (EUF-CMA))}: Our scheme is EUF-CMA secure if no adversary $\mathcal{A}$ possesses a non-negligible advantage in winning the following game.

\textbf{Initial}: A challenger $\mathcal{C}$ initializes the system parameters and master key. Then the parameters is forwarded to $\mathcal{A}$.

\textbf{Attack}: $\mathcal{A}$ proposes various queries adaptively as described in the previous definition. $\mathcal{C}$ is able to answer the queries.

\textbf{Forgery}: $\mathcal{A}$ outputs the forged ciphertext and the identities of a UAV and an SDP controller. It is worth noting that the identities and packet cannot be queried in the attack phase. If the forged ciphertext is recognised as valid, $\mathcal{A}$ wins the game.

\begin{table}
\renewcommand\arraystretch{1.3}
\caption{Notation and Description\label{tab:notation}}
\begin{tabular*}{\linewidth}{p{0.2\linewidth}|p{0.7\linewidth}}
\hline
        Notations & Descriptions \\ \hline
        $ssk$     &  System public key \\
        $spk$     &  System master key\\
        $ID_C$    &  The identity of SDP controller\\
        $PK_C$$/$$SK_C$   &  The public/private key of $ID_C$\\
        $U_i$     &  The $i$-th UAV\\
        $ID_U^i$  &  The identity of $U_i$\\
        $Rep_i$   &  The reputation of $U_i$\\
        $R_h$$/$$R_l$ &  The threshold of reputation\\        
       ($C_i$, $R_i$)     &  The challenge–response pair of $U_i$\\
        $PK_U^i$$/$$SK_U^i$  &  The public/private key of $U_i$\\ 
        $pac$     &  An SPA packet generated by $U_i$\\
        $\sigma$  &  The ciphertext of $pac$ generated by $U_i$\\  \hline
\end{tabular*}
\end{table}

\section{BAZAM: Zero-Trust Authentication Scheme}
The proposed BAZAM is organized into three phases, namely the initialize phase, the registration phase, and the UAV-SDP controller authentication phase.  Table \ref{tab:notation} summarizes the notations used in this paper.

\subsection{Initialization}
In the proposed authentication phase, UAVs are authenticated under the coordination of SDP controller and KGC. Before starting the process, the system is assumed to have been initialized, which means:
\begin{enumerate}
\item{System parameters have been instantiated. Let $\mathbb{G}_1$ be a cyclic additive group generated by $G$, and $\mathbb{G}_2$ be a cyclic multiplicative group. $\mathbb{G}_1$ and $\mathbb{G}_2$ have the same order $q$, where $q$ is a primer number of $k$-bits length. $e:$ $\mathbb{G}_1$ $\times$ $\mathbb{G}_1$$\xrightarrow{}$$\mathbb{G}_2$ is a bilinear map. Then, $H_{1}$, $H_{2}$, and $H_{3}$ denote the secure one-way hash functions with the definitions as follows: $H_{1}$: $\{$0,1$\}$$^{\ast}$$\xrightarrow{}$$\mathbb{G}_1$, $H_{2}$: $\mathbb{G}_2$$\xrightarrow{}$$\{$0,1$\}$$^{l}$, and $H_{3}$: $\{$0,1$\}$$^{l}$$\times$ $\mathbb{G}_2$$\xrightarrow{}$$\mathbb{Z}_p^*$, where $l$ denotes a bit length of a message. Let $En$ and $De$ be the encryption and decryption algorithms of a secure symmetric cipher.}
\item{System public key and master key have been generated. It first chooses randomness $ssk$ $\in \mathbb{Z}_p^*$ to compute $spk$ = $sskG$. It finally returns a public key $spk$ and master key $ssk$. It is noticing that $ssk$ is kept secretly and known only by KGC.}
\end{enumerate}

\begin{figure}[!t]
\centering
\includegraphics[width=\linewidth]{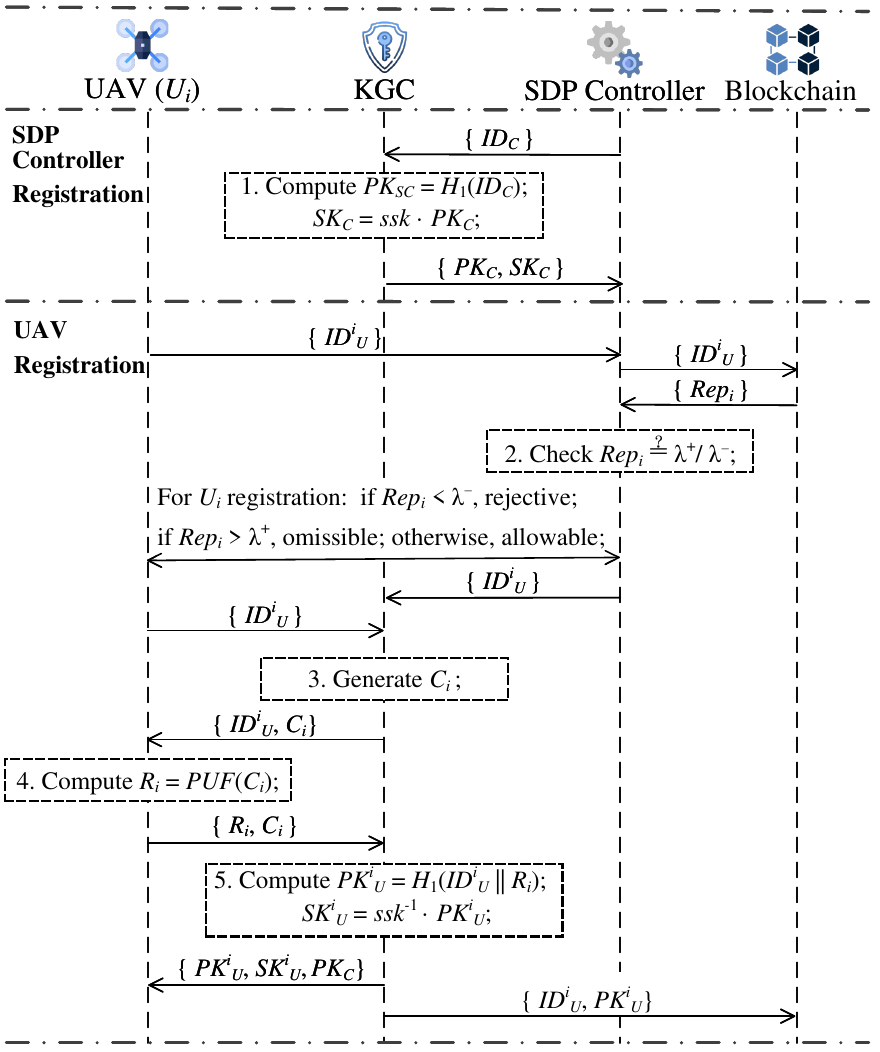}
\caption{The registration process of a UAV and an SDP controller.}
\label{fig:Registration}
\end{figure}

\subsection{SDP Controller Registration}
As illustrated in Fig. \ref{fig:Registration}, the public/private key pair of the SDP controller $ID_{C}$ is generated by KGC in this phase. The public key of $ID_{C}$ generated by KGC is as follows:
\begin{equation*}
PK_{C} = H_1(ID_{C}).
\end{equation*}

The corresponding private key is computed as:
\begin{equation*}
SK_{C}= sskPK_{C}.
\end{equation*}

Then, KGC sends $\{$$PK_{C}$, $PK_{C}$ $\}$ to SDP controller. Note that, the public key of $ID_{C}$ is finally shared with the UAVs that submit the registration request to gateway. The private key is stored secretly by SDP controller.

\subsection{UAV Registration}
In the registration phase, each UAV is equipped with a PUF for generating a response output for a challenge input to it. Considering that a malicious UAV is able to continuously request authentication from the system, it could result in a significant increase in local traffic and lead to device failures. To mitigate the occurrence of such situations, the system incorporates a reputation mechanism in the registration phase. Fig. \ref{fig:Registration} presents the registration process of a UAV. The key steps of a UAV registration process are explained below. 

\begin{enumerate}
\item{When UAV $U_i$ with identity $ID_{U}^i$ enters the network for the first time, it first establishes a registration request with the SDP controller via a secure and private channel.}
\item{Upon receiving the registration request, the SDP controller first retrieves the reputation of the UAV on the blockchain. The SDP controller will initiate a retrospective search from the latest block, proceeding backward until locating the block containing pertinent information about $U_i$. This approach ensures the retrieval of the latest UAV information, including its reputation and public key, etc. For $U_i$ with its reputation $Rep_i$ below a threshold $R_l$, the registration request will be rejected by the SDP controller, and registration services will not be provided. For $U_i$ with $Rep_i$ above a threshold $R_h$, it will receive a notification from the SDP controller indicating no registration is required. Otherwise, the SDP controller sends registration request to the KGC with $ID_{U}^i$ of the $U_i$.}
\item{Upon receiving the valid registration request from $U_i$ with its identity, KGC generates a random challenge $C_i$ and sends it to $U_i$.
Then, a random challenge $C_i$ is generated by KGC. KGC sends $\{$$ID_{U}^i$, $C_i$$\}$ to $U_i$. After receiving the $C_i$, based on which a response, $R_i$, is produced applying the PUF as:
\begin{equation*}
R_i = PUF({C_i}).
\end{equation*}
Then, $U_i$ sends $\{$$C_i$, $R_i$$\}$ to the KGC.
}
\item{When KGC receives $R_i$ from $U_i$, KGC generates the public key $PK_{U}^i$ for $U_i$ as:
\begin{equation*}
PK_{U}^i = H_1({ID_{U}^i} || {R_i}).
\end{equation*}
Then, the corresponding private key is established as:
\begin{equation*}
SK_{U}^i = ssk^{-1}\cdot PK_{U}^i.
\end{equation*}
The KGC sends $PK_{U}^i$, $SK_{U}^i$, and $PK_{C}$ to $U_i$.
}
 \item{The set $\{$$ID_{U}^i$, $PK_{U}^i$, $SK_{U}^i$, $PK_{C}$$\}$ is securely stored in the UAV’s storage. Meantime, the KGC uploads the set $\{$$ID_{U}^i$, $PK_{U}^i$$\}$ on the blockchain. }
\end{enumerate}

It should be noted that in the case of a UAV registering for the first time, its reputation is set to zero and uploaded on the blockchain. For the UAV with prior registration history, its reputation remains unchanged during the registration phase. Therefore, when the registration phase is over, the blockchain contains the tuple $<$$ID_{U}^i$, $PK_{U}^i$, $Rep_i$$>$ for the registered $U_i$.

\subsection{UAV-SDP controller Authentication}
\begin{figure}[!t]
\centering
\includegraphics[width=\linewidth]{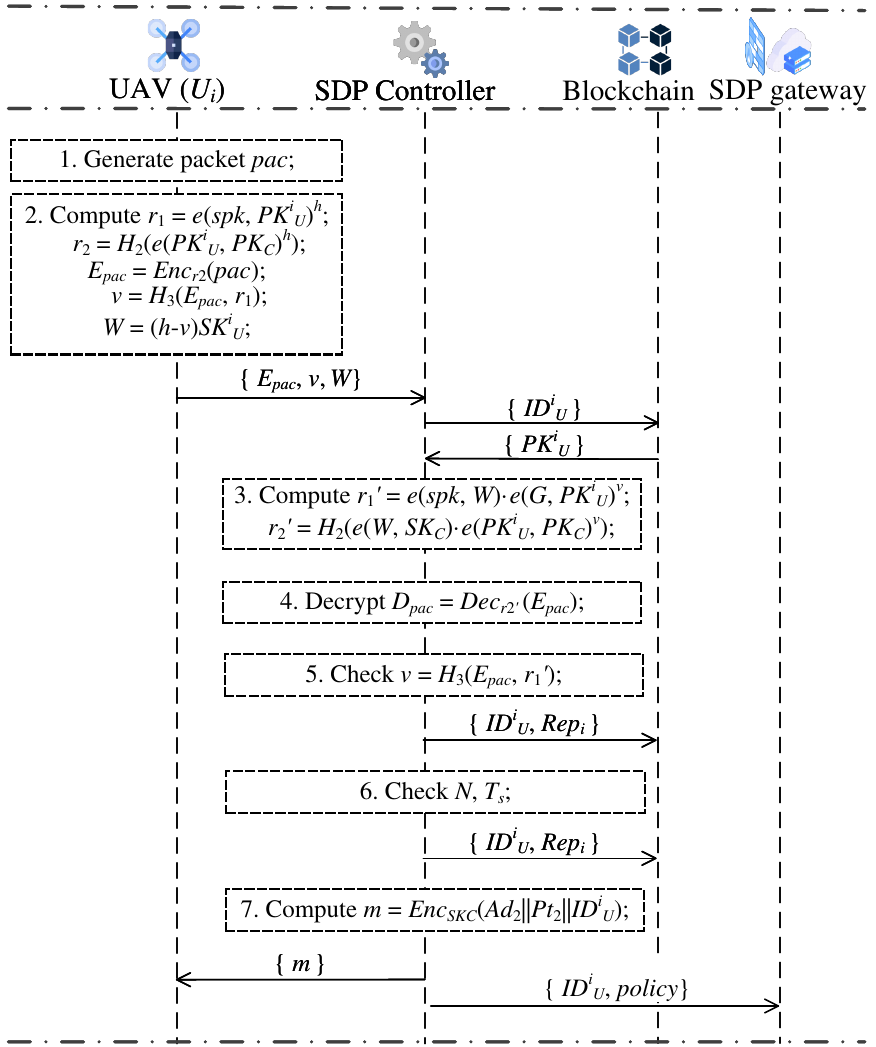}
\caption{The UAV-SDP controller authentication process.}
\label{fig:Authentication}
\end{figure}

Here, we describe our scheme to achieve the authentication between UAV $U_i$ and SDP controller under the zero trust network. Fig. \ref{fig:Authentication} depicts the various authentication phase performed on $U_i$ and SDP controller. This scheme ensures that only registered UAVs can authenticate with SDP controller before they can begin establishing a secure connection with the SDP gateway.

\begin{enumerate}
\item{Prior to the initiation of a connection request by $U_i$, rigorous identity authentication must be conducted through the SDP controller. 
\begin{itemize}
\item[--] First, $U_i$ generates an SPA packet based on its own information and requirements as:
\begin{equation*}
pac = (N||ID_{U}^i||Pwd||Ts||Ver||Ad_1||Pt_1),
\end{equation*}
where $N$ denotes the nonce number, $Pwd$ denotes the password, $Ts$ denotes the timestamp, $Ver$ denotes the protocol version, $Ad_1$ denotes the IP address, and $Pt_1$ denotes the port number.
\item[--] $U_i$ selects a random number $h$$\in \mathbb{Z}_p^*$ to derive $k_1$ and $k_2$ as:
\begin{equation*}
r_1 = e(spk, PK_{U}^i)^h,
\end{equation*}
\begin{equation*}
r_2 = H_2(e(PK_{U}^i, PK_{C})^h).
\end{equation*}
\item[--] Based on the $pac$, $r_1$, and $r_2$, $U_i$ performs the following operations to generate the ciphertext $\sigma$ = ($E_{pac}$, $v$, $W$) as:
\begin{equation*}
E_{pac} = Enc_{r_2}(pac),
\end{equation*}
\begin{equation*}
v = H_3(E_{pac}, r_1),
\end{equation*}
\begin{equation*}
W = (h-v)SK_{U}^i,
\end{equation*}
where $Enc$ is the encryption algorithm of a secure symmetric cipher of length $l$.
\end{itemize}
}

\item{$U_i$ then sends $\sigma$ as an authentication request to the SDP controller.}
\item{After receiving the authentication request from $U_i$, SDP controller executes the following steps to verify the ciphertext:
\begin{itemize}
\item[--] It first retrieves the public key of $U_i$ stored in the blockchain and then calculates
\begin{equation*}
r_1' = e(spk, W)\cdot e(G, PK_{U}^i)^v,
\end{equation*}
\begin{equation*}
r_2' = H_2(e(W, SK_{C})\cdot e(PK_{U}^i, PK_{C})^v).
\end{equation*}
\item[--]The plaintext $pac'$ is derived as:
\begin{equation*}
D_{pac} = Dec_{r_2'}(E_{pac}),
\end{equation*}
where $Dec$ is the corresponding decryption algorithm of a secure symmetric cipher of length $l$.
\item[--] The SDP controller checks the plaintext to ensure the data integrity as:
\begin{equation*}
v = H_3(E_{pac},r_1').
\end{equation*}
\item[--] If the checking is successful, the ciphertext is valid and the $pac'$ is obtained. 
\item[--] Otherwise, the ciphertext is rejected and the authentication is failed. In the event of a failed authentication for $U_i$, the SDP controller will undertake an adjustment of its reputation, decrementing it by one unit. Subsequently, the updated reputation $Rep_i$ of $U_i$, along with its identity, will be encapsulated into a transaction, and  stored in a new block.
\end{itemize}
}
\item{Afterward, the SDP controller initially verifies the correctness of the $Pwd$ of $ID_{U}^i$. Then, it meticulously scrutinizes the presence of $N$ and $Ts$ to ascertain the absence of replay attacks. Similar to the previous step, the reputation of UAV, failing authentication, will decrease and be updated on the blockchain. }
\item{Upon the successful authentication, the SDP controller communicates the IP address $Ad_2$ and port $Pt_2$ of gateway back to $U_i$ as:
\begin{equation*}
m = Enc_{SK_{C}}(Ad_2||Pt_2||ID_{U}^i).
\end{equation*}
It is noteworthy that the reputation of a successfully authenticated $U_i$ will be incremented by one unit and uploaded on the blockchain.}
\item{The SDP controller simultaneously communicates the legitimate UAV identity and the authorization policy to the SDP gateway.}
\item{ $U_i$ sends a connection request and the authorization to SDP gateway with its identity. Since the UAV has been validated by the SDP controller, a secure session is established between $U_i$ and gateway.}

\end{enumerate}

\subsection{Reputation Update}
The proposed BAZAM utilizes the reputation value to indicate the trustworthiness of UAVs' identities. The reputation of a UAV is quantified by a specific numerical value, reflecting the long-term performance of its authentication behavior. A higher value of the UAV corresponds to a greater number of legitimate authentication instances, indicating a higher level of trust. Without loss of generality, the initial reputation value of the UAV is set to zero. While the reputation value undergoes updates, it does not reset or zero out.

Considering the need to adhere to the zero-trust principle of ``never trust, always verify," every UAV is required to undergo authentication before establishing a secure connection with the SDP gateway. The reputation of a UAV evolves with each authentication attempt. Therefore, to simplify the entire authentication process, a comparison between the reputation stored in the blockchain and the predefined thresholds is employed.

For a UAV with a reputation below the threshold $R_l$, indicating multiple authentication failures, the SDP controller will blacklist it. Consequently, the SDP controller will reject its registration request to the KGC, preventing it from obtaining fresh key pair. Similarly, to inspire well-behaved UAVs in terms of authentication and enhance the authentication speed, a threshold $R_h$ is set. This threshold is used to identify high-reputation UAVs, allowing them to bypass the registration process. They can simply initiate access requests to the SDP controller using the previously registered public/private key pairs.

\section{Security Analysis}
This section demonstrates the correctness of the BAZAM and security analysis. The analysis results indicate that the proposed scheme is secure against various common attacks.

\textit{Theorem 1 (Confidentiality)}: \textit{Assume that the MDBDH assumption holds, then BAZAM is secure against adaptive chosen-ciphertext attacks.}

\textit{Proof}: Supposing that an adversary $\mathcal{A}$ is able to break the confidentiality of BAZAM with a non-negligible probability $\epsilon$, then a challenger $\mathcal{C}$ has the ability to solve the MDBDH problem with advantage $\epsilon'$ by interacting with $\mathcal{A}$. $\mathcal{C}$ maintains the lists: $L_i$($i$ = 1, 2, 3) for query and response pairs to random oracle $H_i$.

\textbf{Initial}: $\mathcal{C}$ submits the system parameters to $\mathcal{A}$ with $spk$ = $cG$, where $ssk$ is the master key that is unknown to $\mathcal{C}$.

\textbf{Phase 1}: $\mathcal{A}$ can issue some queries and $\mathcal{C}$ answers these queries as follows:
\begin{itemize}
\item{$H_1$ query}: $\mathcal{C}$ keeps a list $L_1$ of tuples ($ID_i$, $PK_i$, $s$). Whenever $\mathcal{A}$ makes a query on $H_1(ID_i)$, if the request has been previously queried, the corresponding value in $L_1$ will be returned. Otherwise, $\mathcal{C}$ chooses $s$$\in\mathbb{Z}_p^*$ and computes $PK_i$ = $sG$. Afterwards, $\left(ID_i,PK_i,s\right)$ is stored in $L_1$ by $\mathcal{C}$ and $PK_i$ is sent to $\mathcal{A}$. $\mathcal{C}$  chooses a extra query with random $ID_n$ and responses $H_1(ID_n)$ = $bG$.
\item{$H_2$ query}: $\mathcal{C}$ keeps a list $L_2$ of tuples ($\tau$, $r_{2}$). Whenever $\mathcal{A}$ makes a query on $H_2(\tau)$, if the request has been previously queried, the corresponding value in $L_2$ will be returned. Otherwise, $\mathcal{C}$ chooses $r_{2}$$\in$$\{$0,1$\}$$^{l}$ and returns it.
\item{$H_3$ query}: $\mathcal{C}$ keeps a list $L_3$ of tuples $\left( E_{pac},r_{1},v \right)$. Whenever $\mathcal{A}$ makes a query on $H_3\left(E_{pac}, r_{1}\right)$, if the request has been previously queried, the corresponding value in $L_3$ will be returned. Otherwise, $\mathcal{C}$ chooses $v$$\in \mathbb{Z}_p^*$ and returns it.
\item{Key extract query}: Suppose that $\mathcal{A}$ has made the $H_1$ query with the UAV $ID_{U}$ and SDP controller $ID_{C}$, where $ID_{U}$ is the aggregate identity with challenge-response of UAV. Note that if these two identities equal to $ID_n$, $\mathcal{C}$ cannot calculates the private keys and declares it a failure. Then $\mathcal{C}$ calculates $SK_{U}$ = $s$$c^{-1}G$ and $SK_{C}$ = $s$$cG$. Finally, $\mathcal{C}$ returns $SK_{U}$ and $SK_{C}$ to $\mathcal{A}$.
\item{Signcrypt query}: Once upon receiving the query with $ID_{U}^j$, $ID_{C}$, and a packet $pac$, $\mathcal{C}$ calculates $SK_{U}$ if $ID_{U}^j$ $\neq$ $ID_n$. Then $pac$ is encrypted and signed to $\mathcal{A}$. Otherwise, $\mathcal{C}$ can calculate $SK_{C}$. Then $\mathcal{C}$ randomly selects $v$$\in \mathbb{Z}_p^*$ and $W$$\in \mathbb{G}_1$ to derive $r_{1}$, $r_{2}$, and $E$. If a tuple $\left(E_{pac},r_{1},v'\right)$ and $v'$ $\neq$ $v$ exists in $L_3$, then $\mathcal{C}$ selects $v$ and $W$ again until no such a tuple in $L_3$.
\item{Unsigncrypt query}: Once upon receiving the query with $ID_{U}^j$, $ID_{C}$, and a ciphertext $D_{pac}$, $\mathcal{C}$ returns response to $\mathcal{A}$. For the case $ID_{C}$ = $ID_n$, $\mathcal{C}$ returns invalid ciphertext to $\mathcal{A}$. However, if $\mathcal{A}$ has got an answer $v$ from $\mathcal{C}$ by making $H_3$ query, then $\mathcal{C}$ fails. Otherwise, $\mathcal{C}$ can calculate $SK_{C}$ and $r_{1}$ to check the existence of the tuple $\left(E_{pac},r_{1},v\right)$ in $L_3$. If so, $\mathcal{C}$ recovers $v$ and calculates $\tau$. If the $L_2$ does not exist the tuple ($\tau$, $\cdot$), $\mathcal{C}$ randomly selects $\tau'$ and $r_2$ and stores them in a tuple. Afterwards, $\mathcal{C}$ derives $D_{pac}$ and sends it to $\mathcal{A}$. 
\end{itemize}

\textbf{Challenge}: $\mathcal{A}$ outputs two plaintexts $pac_1$ and $pac_2$ , the identities $ID_{U}$ and $ID_{C}$. If $ID_{C}$ is queried in the key extract query, $\mathcal{C}$ fails. Otherwise, $\mathcal{C}$ randomly selects $\beta$$\in$$\{$0,1$\}$, $v$$\in \mathbb{Z}_p^*$. Finally, $\mathcal{C}$ sets $W^*$ = $aG$ and returns ${pac}_{\beta}$'s challenge ciphertext $\sigma^*$ = $(E_{{pac}_\beta}^*,v^*,W^*)$ to $\mathcal{A}$.

\textbf{Phase 2}: After receiving the $\sigma^*$, $\mathcal{A}$ is allowed to make queries in the phase 1. However, it cannot make the key extract query with $ID_{C}$ and the make the unsigncrypt query with $ID_{U}$ and $ID_{C}$ to $\sigma^*$.

\textbf{Guess}: $\mathcal{A}$ outputs $\beta'$ as its answer. If $\beta'$ = $\beta$, $\mathcal{C}$ outputs $R$ = $e(G,G)^{abc}$ as the solution to the MDBDH problem.

Suppose that $\mathcal{A}$ can make maximum $q_{H_i}$ times $H_i$ queries ($i$ = 1, 2, 3), $q_{ke}$ times key extract queries, $q_{H_3}$ times signcrypt queries, $q_{us}$ times unsigncrypt queries. Since $\sigma^*$ is generated randomly by $\mathcal{C}$, $\mathcal{A}$ cannot provide any advantage. However, $\mathcal{A}$ has a non-negligible advantage $\epsilon$ to recognize the challenge is incorrect. For $\mathcal{C}$, the probability that it rejects a valid ciphertext in unsigncrypt query is 1$-$$(\frac{2^k-1}{2^k})$$^{q_{us}}$$\leqslant$$\frac{q_{us}}{2^k}$. Then, $\mathcal{C}$ would continue the simulation in the following situations:

1) $S$1: $\mathcal{A}$ does not make an key extract query with the $ID_{n}$ in the challenge phase. Then, Pr[$S$1] = $\frac{{q_{H_1}}-{q_k}}{q_{H_1}}$.

2) $S$2: The identities chosen by $\mathcal{A}$ satisfy $ID_{U}$ $\neq$ $ID_{n}$ and $ID_{C}$ = $ID_{n}$. Then, Pr[$S$2] = $\frac{1}{{q_{H_1}}-{q_k}}$.

Therefore, the probability that $\mathcal{C}$ can get the answer of MDBDH problem given by $\mathcal{A}$ is
\begin{equation*}
Pr[S1\wedge S2]=(\frac{{q_{H_1}}-{q_k}}{q_{H_1}})(\frac{1}{{q_{H_1}}-{q_k}})=\frac{1}{q_{H_1}},
\end{equation*}
and $\mathcal{A}$ has a non-negligible advantage $\epsilon$, the probability $\mathcal{C}$ can solve the MDBDH problem is 
\begin{equation*}
\begin{aligned}
& Pr\left[\mathcal{C}(G,aG,bG,cG,c^{-1}G)|a,b,c\in \mathbb{Z}_p^* \right]\\
& \geqslant \left[\frac{\epsilon+1}{2}\left( 1- \frac{q_{us}}{2^k} \right) - \frac{1}{2}\right]\frac{1}{q_{H_1}} = \frac{{\epsilon(2^k-q_{us})}-q_{us}}{q_{H_1}2^{k+1}}.
\end{aligned}
\end{equation*}

Hence, $\mathcal{C}$ can solve the MDBDH problem with a non-negligible probability. However, this is contrary to the Definition 2. Hence, the proposed BAZAM realizes confidentiality.

\textit{Theorem 2 (Unforgeability)}: \textit{Assume that the MBDH assumption holds, then BAZAM is existential unforgeability against adaptive chosen-message attacks.}

\textit{Proof}: Suppose that an adversary $\mathcal{A}$ is able to break the unforgeability of BAZAM with a non-negligible probability $\epsilon$, then a challenger $\mathcal{C}$ has the ability to solve the MBDH problem in an expected time by interacting with $\mathcal{A}$. $\mathcal{C}$ maintains the lists: $L_i$($i$ = 1, 2, 3) for query and response pairs to random oracle $H_i$.

\textbf{Initial}: $\mathcal{C}$ submits the system parameters to $\mathcal{A}$ with $spk$ = $sskG$, where $ssk$ is the master key that is unknown to $\mathcal{C}$.

\textbf{Attack}: $\mathcal{A}$ is able to adaptively query for various oracles. $\mathcal{C}$ answers the queries as in Theorem 1.

\textbf{Forgery}: $\mathcal{A}$ outputs the forged the packet $pac$ and the identity $ID_{U}$. If a valid tuple $\left(pac,ID_{U}\right)$ is output, $\mathcal{C}$ employs the forking lemma in the literature \cite{ana1}: If $\mathcal{A}$ can output a valid signature tuple within time $t$ with a advantage $\epsilon$ $\geqslant$ $\frac{10(q_{us})({q_{us}}+q_{H_3}){q_{H_1}}}{2^k-1}$, the expected time that $\mathcal{C}$ can get two valid signature tuples is less than 120686$\frac{2^k{q_{H_3}}q_{H_1}t}{\epsilon (2^k-1)}$.

After obtaining two signature tuples $\left((pac,ID_{U}),v,W\right)$ and $\left((pac,ID_{U}),v',W'\right)$, where $v$ $\neq$ $v'$, another machine $\mathcal{C'}$ has control over the machine obtained from $\mathcal{A}$ replacing interaction with the UAV by simulation and gains the two valid signature tuples. $\mathcal{C'}$ calculates $bc^{-1}G$ = $(v'-v)^{-1}$$(W-W')$. Then $\mathcal{C'}$ can derive $e(G,G)^{{abc}^{-1}}$ = $e(aG,bc^{-1})G$.

Hence, $\mathcal{C}$ can solve the MBDH problem with a non-negligible probability. However, this is contrary to the Definition 1. Therefore, the proposed BAZAM realizes unforgeability.

\textit{Resistance of replay attack:} In this attack, an adversary performs a replay attack by replaying formerly received legal messages. Nevertheless, since the SPA packet generated when the UAV initiates an authentication request includes nonce number $N$ and timestamp $T_s$, the SDP controller will assess whether the packet has been replayed based on $T_s$ of the current message. Therefore, the BAZAM scheme can resist reply attacks.

\textit{Perfect forward secrecy:} An adversary has the ability to successfully steal the private/public key pairs of the UAVs. It cannot obtain the keys of any previous authentication due to the response $R_i$ and $ssk$ corresponding to generating these keys. Consequently, the BAZAM scheme ensures perfect forward secrecy.

\textit{Perfect backward secrecy:}
In the proposed scheme, even if an adversary manages to guess the current private key, it will not compromise the security of the next authentication. This is because once the reputation value falls below the threshold $R_h$, the UAV is required to register with the KGC each time it submits an access request. Furthermore, even if the private key is obtained, it is difficult for adversary to compute $k_2$ and decrypt the plaintext.

\textit{UAV physical capture attack:}
In the UAV registration process, suppose a UAV $U_j$ is physically captured by an adversary, and then its identity is extracted in the memory. However, the adversary cannot derive the private/public key pair due to the unknown master key and a random nonce $R_j$. Moreover, there are distinct and independent for the deployed UAV since PUF challenge and response pair ($C_j$,$R_j$) are randomly generated. Hence, the revealed data does not help in obtaining the key pair. Therefore, the BAZAM scheme can prevent UAV physical capture attacks due to its reliance on the intrinsic physical variations within the integrated circuit chip, which determines the outputs of the PUF challenge and response pairs.

\textit{UAV traceability:}
After successful registration, the SDP controller stores the true identities of the UAVs in the blockchain database. During the authentication process, the SDP controller also promptly updates UAVs' identities and reputation values. If a UAV fails authentication multiple times, the SDP controller can query its reputation value in the blockchain based on its identity and decide whether to accept subsequent authentication requests from the UAV.

\section{Performance Analysis}

In this section, we carry out experiments to assess the performance of BAZAM, including computation overhead, communication overhead, and storage overhead.

\subsection{Experimental Settings}
This section presents the performance timings of various cryptographic primitives, leveraging the widely-acknowledged ``Pairing-Based Cryptography Library (PBC)" with the Type A curve. The PBC is a C based programming software library, designed to perform mathematical operations for pairing-based encryption systems. We evaluate the computation time for various cryptographic primitives via experiments. Let $T_{ad1}$, $T_{mu1}$, $T_{mu2}$, $T_{ex}$, $T_{bp}$, and $T_h$ denote the notations of point addition in $\mathbb{G}_1$, point multiplication in $\mathbb{G}_1$, point multiplication in $\mathbb{G}_2$, exponentiation operations in $\mathbb{G}_2$, bilinear pairing, and hash operations, respectively. 

The entire experiments have been done over version of PBC 0.5.14 in a virtual machine of Ubuntu 16.04 hosted on VMware Workstation 15 Pro. In this platform, we executed each cryptographic primitive for 1000 times and considered the average run time for each cryptographic primitives from these 1000 runs. Then we have $T_{ad1}$ $\approx$ 0.005 \textit{ms}, $T_{mu1}$ $\approx$ 0.926 \textit{ms}, $T_{mu2}$ $\approx$ 0.003 \textit{ms}, $T_{ex}$ $\approx$ 0.098 \textit{ms}, $T_{bp}$ $\approx$ 0.757 \textit{ms}, and $T_h$ $\approx$ 0.003 \textit{ms}.

\subsection{Computation Overhead}

\begin{figure}[!t]
\centering
\includegraphics[width=0.7\linewidth]{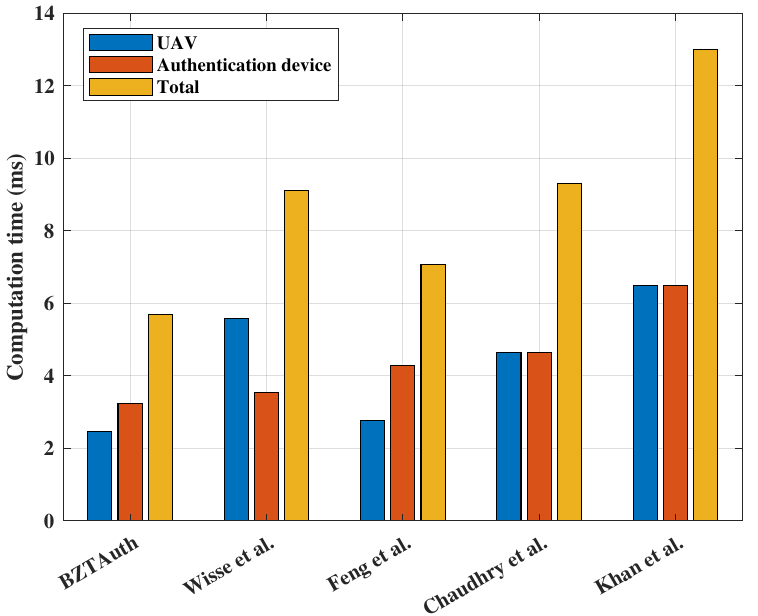}
\caption{Computation cost of different entities in various schemes (in ms).}
\label{fig:Compu}
\end{figure}

We have compared the proposed scheme with state-of-the-art schemes of Wisse \textit{et al}.\cite{comp1}, Feng \textit{et al}.\cite{comp2}, Chaudhry \textit{et al}.\cite{comp3}, and Khan \textit{et al}.\cite{comp4}. The specific computational cost of each entity and total computational cost in different schemes are shown in Table \ref{tab:compu}. Considering the authentication process, it highlights the computational cost of BAZAM and other schemes. In BAZAM, UAV and SDP controller need to perform 3$T_{bp}$ + 2$T_{h}$ + 2$T_{ex}$ and 4$T_{bp}$ + 2$T_{h}$ + 2$T_{ex}$ operations, respectively, which yields a total cost of 5.703ms. In BAZAM, the UAV sends its resource access request packet to the SDP controller for cryptographic processing and verification. Therefore, the most time-consuming phase is the encryption and signing processing of packets by UAV. In Wisse \textit{et al}.'s \cite{comp1} scheme, the authenticity of broadcast messages emitted by UAV is verified by an observer. During the identity authentication process, both the signatures generated by the UAV and the ciphertext recovered by observers require 13 $T_{ex}$ operations. In Feng \textit{et al}.'s\cite{comp2} scheme, two pairs of elliptic curve encryption/decryption algorithms are employed in the authentication process. In Chaudhry \textit{et al}.'s\cite{comp3} and Khan \textit{et al}.'s\cite{comp4} schemes, the UAV is required to authenticate with another UAV by sharing the secret keys. Then more $T_{mu1}$ operations are needed than our proposed scheme. It is worth noting that since zero-trust networks are required to continuously authenticate UAVs, the comparison overhead only considers a single authentication of the UAV. From the Fig. \ref{fig:Compu}, it indicates that BAZAM is more computational cost-effective than the other schemes.

\begin{table*}
\renewcommand\arraystretch{1.4}
    \centering
    \caption{Comparison between the proposed scheme and existing schemes of computation overhead\label{tab:compu}}
\begin{tabular}{lll}
\hline
Scheme & UAV & Authentication Device      \\ \hline
BAZAM   & 3$T_{bp}$ + 2$T_{h}$ + 2$T_{ex}$ $\approx$ 2.473 \textit{ms} & 4$T_{bp}$ + 2$T_{h}$ + 2$T_{ex}$ $\approx$ 3.23 \textit{ms} \\
Wisse \textit{et al}.\cite{comp1}     &  2$T_{bp}$ + 2$T_{h}$ +  13$T_{ex}$ + 3$T_{mu1}$ + 2$T_{mu2}$ $\approx$ 5.578 \textit{ms} &  3$T_{bp}$ + 13$T_{ex}$ $\approx$ 3.545 \textit{ms} \\
Feng \textit{et al}.\cite{comp2}     & 3$T_{mu1}$ + $T_{ad1}$ $\approx$ 2.783 \textit{ms} &  3$T_{mu1}$ + $T_{ad1}$ + 2$T_{bp}$ + $T_{h}$ $\approx$ 4.3 \textit{ms}   \\
Chaudhry \textit{et al}.\cite{comp3}    &  5$T_{mu1}$ + 2$T_{ad1}$ + 3$T_{h}$ $\approx$ 4.649 \textit{ms}  & 5$T_{mu1}$ + 2$T_{ad1}$ + 3$T_{h}$ $\approx$ 4.649 \textit{ms} \\
Khan \textit{et al}.\cite{comp4}     &  7$T_{mu1}$ + $T_{ad1}$ + 3$T_{h}$ $\approx$ 6.496 \textit{ms} & 7$T_{mu1}$ + $T_{ad1}$ + 3$T_{h}$ $\approx$ 6.496 \textit{ms}  \\
\hline
\end{tabular}
\end{table*}

\subsection{Communication Overhead}

\begin{table*}
\renewcommand\arraystretch{1.4}
    \centering
    \caption{Comparison between the proposed scheme and existing schemes of communication overhead\label{tab:commu}}
\begin{tabular}{lll}
\hline
Scheme & UAV & Authentication Device     \\ \hline
BAZAM   & $|\mathbb{G}_1|$ + $|\mathbb{Z}_p|$ + $|H|$ & 4$|ID|$ + $|m|$ + 3$|\mathbb{Z}_p|$ \\
Wisse \textit{et al}.\cite{comp1}     &  7$|\mathbb{G}_1|$ + 3$|\mathbb{G}_2|$  + $|m|$ & - \ \\
 Feng \textit{et al}.\cite{comp2}     & 2$|\mathbb{G}_1|$ + 2$|\mathbb{Z}_p|$ + 2$|m|$ + $|ID|$ + $|E|$ & 2$|\mathbb{G}_1|$ + $|\mathbb{Z}_p|$    \\
Chaudhry \textit{et al}.\cite{comp3}    &  $|\mathbb{G}_1|$ + $|\mathbb{Z}_p|$ + $|ID|$ + 2$|H|$ + 2$|T|$& $|\mathbb{G}_1|$ + $|\mathbb{Z}_p|$ + $|ID|$ + $|H|$ + $|T|$ \\
Khan \textit{et al}.\cite{comp4}     & $|\mathbb{G}_1|$ + $|\mathbb{Z}_p|$ + $|ID|$ + $|H|$ + $|T|$  & $|\mathbb{G}_1|$ + $|\mathbb{Z}_p|$ + $|ID|$ + $|H|$ + $|T|$ \\
\hline
\end{tabular}
\end{table*}

\begin{figure}[!t]
\centering
\includegraphics[width=0.7\linewidth]{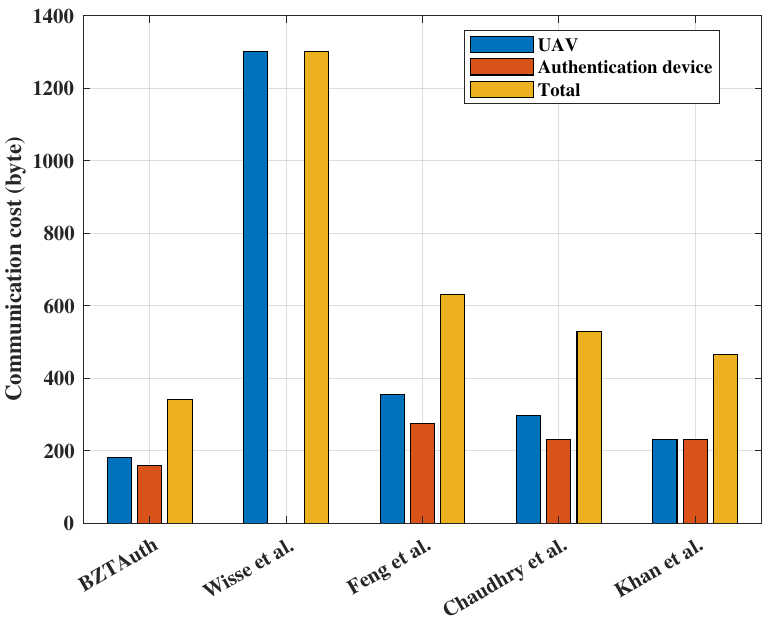}
\caption{Communication cost of different entities in various schemes (in byte).}
\label{fig:Commu}
\end{figure}

As the length of the message varies depending on the task, only the authentication-related communication overloads are computed. As shown in Table \ref{tab:commu}, to evaluate communication overheads, we consider that the sizes of the element in $\mathbb{G}_1$, element in $\mathbb{G}_2$, element in $\mathbb{Z}_p$, identity, digest of hash function, plaintext/ciphertext, and timestamp are $|\mathbb{G}_1|$ = 128 bytes, $|\mathbb{G}_2|$ = 128 bytes, $|\mathbb{Z}_p|$ = 20 bytes, $|ID|$ = 20 bytes, $|H|$ = 32 bytes, $|m|$/$|E|$ = 20 bytes, and $|T|$ = 20 bytes. ``-" indicates that the entity has no associated overhead. In BAZAM, the communication overhead mainly stems from querying the reputation of the UAV and verifying the correctness of the packet information sent by the UAV to determine its identity. The communication overhead in Wisse \textit{et al}.'s \cite{comp1} scheme is totally produced from the UAV. The UAV is required to send the signature and message to the observer. On the receiving side, the observer extracts the message and parses the signature to directly authenticate the UAV and verify its contents. In Feng \textit{et al}.'s\cite{comp2} scheme, the UAV sends ciphertext to the authentication device. If parsing of the elements of the ciphertext verifies a successful formula, then the UAV will be authenticated and the connection will be established. In Chaudhry \textit{et al}.'s\cite{comp3} scheme, the UAV sends information including identity and timestamp, etc., for mutual authentication. Similarly, UAV sends a tuple consisting its identity and timestamp to the authentication device in Khan \textit{et al}.'s\cite{comp4} scheme. From the Fig. \ref{fig:Commu}, the communication cost of BAZAM is similar or less to those of some related schemes. In addition, in the zero-trust network, the SDP gateway defaults to rejecting all access requests. Only after the UAV is authenticated, the SDP gateway establishes a connection with the UAV, thus minimizing  the attack surface and greatly reducing the network security risk. Therefore, BAZAM provides better security and functionalities than related schemes.

\begin{figure}[!t]
\centering
\includegraphics[width=0.7\linewidth]{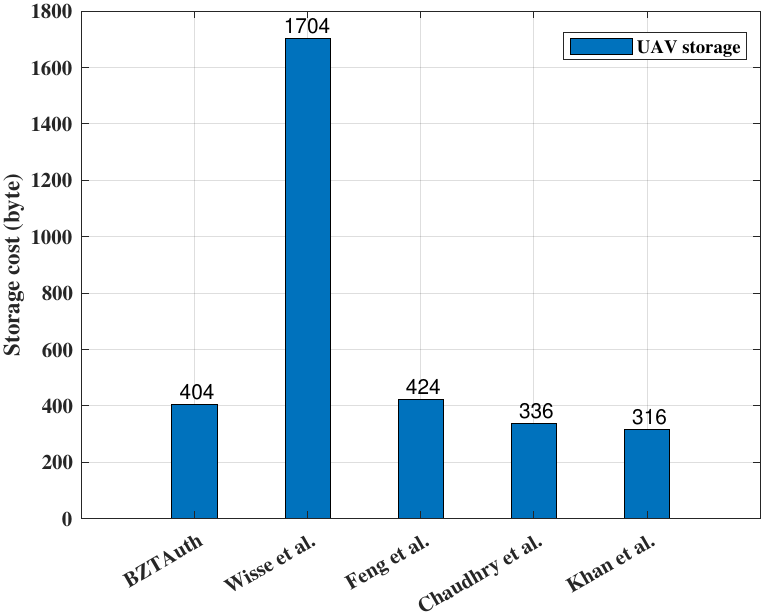}
\caption{Storage cost of various schemes (in byte).}
\label{fig:Store1}
\end{figure}

\subsection{Storage Overhead}
\begin{figure}[!t]
\centering
\includegraphics[width=0.7\linewidth]{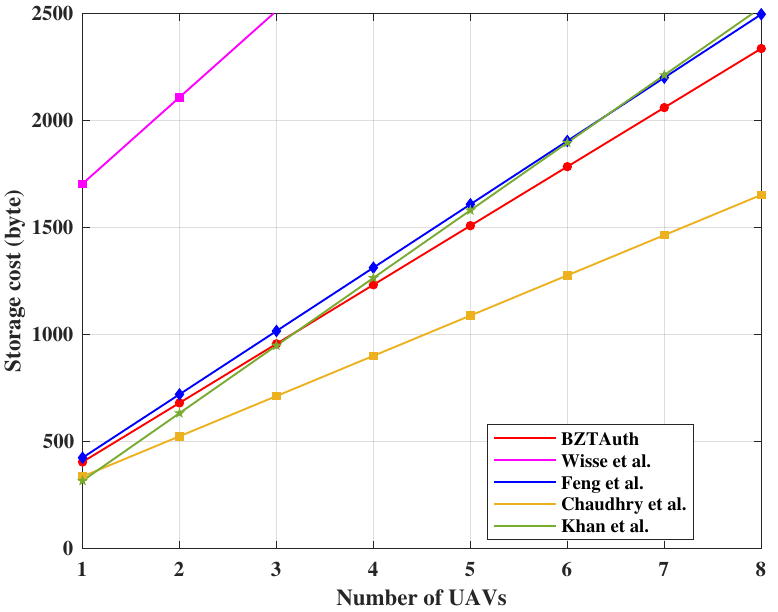}
\caption{Comparison of storage cost versus the UAV number (in byte).}
\label{fig:Store2}
\end{figure}

We assess the additional storage overhead encompassing certain system security parameters and essential cryptographic keys. The storage overhead comparison is shown in Fig. \ref{fig:Store1}. The UAV in BAZAM needs to store identity $ID_{U}^i$, public/ private key pair ($PK_{U}^i$, $SK_{U}^i$), and public key $PK_{C}$ of SDP controller in the memory. It is worthy noting that the challenge is input to generate its public/ private key pair without additional storage. In addition, the reputation value is stored on the blockchain, reducing the UAV's storage overhead. The UAV in Wisse \textit{et al}.'s \cite{comp1} scheme has a storage cost of public group key (1300 bytes) and private key (404 bytes) since the parameters are complex. Similarly, the UAV in Feng \textit{et al}.'s\cite{comp2} scheme requires to store its identity, public/ private key pair, and public key of authentication device. Besides, the challenge $C_i$ (20 bytes) also is needed to be stored in the UAV’s memory. In Chaudhry \textit{et al}.'s\cite{comp3} scheme, the UAV stores its own identity, public and private keys, and its certificate (20 bytes), as well as the identity and public key of the ground station server in the memory. In Khan \textit{et al}.'s\cite{comp4} scheme, the public/ private key pair, identity, certificate (128 bytes, which differs from that of Chaudhry \textit{et al}.'s\cite{comp3} scheme), and signature (20 bytes) are stored in the UAV's memory for authentication. The relationship between the storage overhead and the UAV number is shown in Fig. \ref{fig:Store2}. We observe that the storage overhead increases with the UAV number increasing. As one can observe, the storage overhead of the Wisse \textit{et al}.'s \cite{comp1} scheme is much higher than that of the other schemes. The storage overhead of BAZAM is not significantly different from that of the other schemes. However, with the increasing number of authenticated UAVs, due to the necessity of regenerating parameters for each UAV, the storage costs of the Feng \textit{et al}.'s\cite{comp2} and Khan \textit{et al}.'s\cite{comp4} schemes gradually exceed those of BAZAM. It is worth noting that the storage cost in Chaudhry \textit{et al}.'s\cite{comp3} scheme remains the smallest. This is becaused that the private key of UAV in \cite{comp3} is a random number (20 bytes). However, it is generated based on the identity and challenge value of UAV in BAZAM. This approach aligns better with the characteristics of UAVs and provides higher security than \cite{comp3}. Additionly, although BAZAM has more storage overhead compared to Chaudhry \textit{et al}.'s\cite{comp3} scheme, the communication and computation overhead in it are much higher compared to ours. Hence, in terms of the computation capabilities of UAVs, and storage and communication overheads, BAZAM demonstrates strong competitiveness than the existing authentication schemes for UAVs.

\section{Conclusion}

In this article, we have investigated the identity authentication in  multi-UAV wireless networks and present a blockchain-assisted zero-trust authentication scheme, called BAZAM. The introduction of zero-trust model emphasizes authentication and authorization at every step in the network. The blockchain is integrated to build mutual trust between different entities of the zero-trust network and to protect the UAV information from tampering. This scheme also authenticates the UAV by encrypting and signing its  access request. Once authenticated, its request can be processed correspondingly. Moreover, BAZAM shows good performance in comparison with the other existing UAV authentication schemes in terms of computation, communication, and storage overheads. The future work could be to consider recording all the authentication processes of UAVs in the distributed blockchain, ensuring the safety and efficiency of authentication while realizing traceability and full transparency of UAV authentication.


\begin{thebibliography}{1}
\bibliographystyle{IEEEtran}

\bibitem{int1}
X. Yuan, Y. Hu, J. Zhang, and A. Schmeink, ``Joint user scheduling and UAV trajectory design on completion time minimization for UAV-aided data collection,'' \textit{IEEE Trans. Wireless Commun.}, vol. 22, no. 6, pp. 3884--3898, June 2023.
\bibitem{int2}
X. Yuan, Y. Hu, and A. Schmeink, ``Joint design of UAV trajectory and directional antenna orientation in UAV-enabled wireless power transfer networks,'' \textit{IEEE J. Sel. Areas Commun.}, vol. 39, no. 10, pp. 3081--3096, Oct. 2021.
\bibitem{int3}
Q. Wu, W. Mei and R. Zhang, ``Safeguarding wireless network with UAVs: A physical layer security perspective,'' \textit{IEEE Wireless Commun.}, vol. 26, no. 5, pp. 12--18, Oct. 2019.
\bibitem{int4}
V. Hassija et al., ``Fast, reliable, and secure drone communication: A comprehensive survey,'' \textit{IEEE Commun. Surveys Tuts.}, vol. 23, no. 4, pp. 2802--2832, 4th Quart., 2021.
\bibitem{int5}
X. Lu et al., ``Reinforcement learning-based physical cross-layer security and privacy in 6G,'' \textit{IEEE Commun. Surveys Tuts.}, vol. 25, no. 1, pp. 425--466, 1st Quart., 2023.
\bibitem{int6}
M. Adil, M. A. Jan, Y. Liu, H. Abulkasim, A. Farouk, and H. Song, ``A systematic survey: Security threats to UAV-aided IoT applications, taxonomy, current challenges and requirements With future research directions,'' \textit{IEEE Trans. Intell. Transp. Syst.}, vol. 24, no. 2, pp. 1437--1455, Feb. 2023.
\bibitem{int7}
X. Ren et al., ``A novel access and handover authentication scheme in UAV-aided satellite-terrestrial integration networks enabling 5G,'' \textit{IEEE Trans. Netw. Service Manag.}, vol. 20, no. 3, pp. 3880--3899, Sept. 2023.
\bibitem{int8}
G. Bansal, V. Chamola, N. Ansari, and B. Sikdar, ``Scalable topologies for time-optimal authentication of UAV swarms,'' \textit{IEEE Netw.}, vol. 36, no. 6, pp. 126--132, Dec. 2022.
\bibitem{int9}
H. Wang, X. Wang, H. Fang, and L. Hanzo, ``GALAMC: Guaranteed authentication level at minimized complexity relying on intelligent collaboration,'' \textit{IEEE Trans. Commun.}, vol. 71, no. 5, pp. 2916--2930, May 2023.
\bibitem{int10}
Z. Liu et al., ``Lightweight trustworthy message exchange in unmanned aerial vehicle networks,'' \textit{IEEE Trans. Intell. Transp. Syst.}, vol. 24, no. 2, pp. 2144--2157, Feb. 2023.
\bibitem{int11}
Y. Liu et al., ``Secure and efficient online fingerprint authentication scheme based On cloud computing,'' \textit{IEEE Trans. Cloud Comput.}, vol. 11, no. 1, pp. 564--578, Mar. 2023.
\bibitem{int12}
S. Rose, O. Borchert, S. Mitchell, and S. Connelly, ``Zero trust architecture", 2020.
\bibitem{int13}
S. Hong, L. Xu, J. Huang, H. Li, H. Hu, and G. Gu, ``SysFlow: Toward a programmable zero trust framework for system security,'' \textit{IEEE Trans. Inf. Forensics Security}, vol. 18, pp. 2794--2809, 2023.
\bibitem{int14}
X. Chen, W. Feng, N. Ge, and Y. Zhang, ``Zero trust architecture for 6G security,'' \textit{IEEE Netw.}, early access, Oct. 20, 2023, doi: 10.1109/MNET.2023.3326356.
\bibitem{int15}
Y. Liu et al., ``Secure and scalable cross-domain data sharing in zero-trust cloud-edge-end environment based on sharding blockchain,'' \textit{IEEE Trans. Dependable Secure Comput.}, early access, Sept. 11, 2024, doi: 10.1109/TDSC.2023.3313799.
\bibitem{int16}
M. Tsai, S. Lee, and S. W. Shieh, ``Strategy for implementing of zero trust architecture,'' \textit{IEEE Trans. Rel.}, vol. 73, no. 1, pp. 93--100, Mar. 2024.
\bibitem{rel1}
L. Gupta, R. Jain, and G. Vaszkun, ``Survey of important issues in UAV communication networks,'' \textit{IEEE Commun. Surveys Tuts.}, vol. 18, no. 2, pp. 1123--1152, 2nd Quart., 2016.
\bibitem{rel2}
Y. Bai, H. Zhao, X. Zhang, Z. Chang, R. Jäntti, and K. Yang, ``Toward autonomous multi-UAV wireless network: A survey of reinforcement learning-based approaches,'' \textit{IEEE Commun. Surveys Tuts.}, vol. 25, no. 4, pp. 3038--3067, 4th Quart., 2023.
\bibitem{rel3}
R. T. Tiburski et al., ``Lightweight security architecture based on embedded virtualization and trust mechanisms for IoT edge devices,'' \textit{IEEE Commun. Mag.}, vol. 57, no. 2, pp. 67--73, Feb. 2019.
\bibitem{adrel1}
J. Wang, Z. Jiao, J. Chen, X. Hou, T. Yang, and D. Lan, ``Blockchain-aided secure access control for UAV computing networks,'' \textit{IEEE Trans. Network Sci. Eng.}, doi: 10.1109/TNSE.2023.3324639.
\bibitem{rel4}
Y. Zhou et al., ``Secure communications for UAV-enabled mobile edge computing systems,'' \textit{IEEE Trans. Commun.}, vol. 68, no. 1, pp. 376--388, Jan. 2020.
\bibitem{rel5}
G. Liu et al., ``Softwarized IoT network immunity against eavesdropping with programmable data planes,'' \textit{IEEE Internet Things J.}, vol. 8, no. 8, pp. 6578--6590, Apr. 2021.
\bibitem{rel6}
M. Adil, M. A. Jan, Y. Liu, H. Abulkasim, A. Farouk, and H. Song, ``A systematic survey: Security threats to UAV-aided IoT applications, taxonomy, current challenges and requirements with future research directions,'' \textit{IEEE Trans. Intell. Transp. Syst.}, vol. 24, no. 2, pp. 1437--1455, Feb. 2023.
\bibitem{rel7}
Y. Tan, J. Liu, and N. Kato, ``Blockchain-based lightweight authentication for resilient UAV communications: Architecture, scheme, and future directions,'' \textit{IEEE Wireless Commun.}, vol. 29, no. 3, pp. 24--31, June 2022.
\bibitem{rel8}
Y. Tan, J. Wang, J. Liu, and N. Kato, ``Blockchain-assisted distributed and lightweight authentication service for industrial unmanned aerial vehicles,'' \textit{IEEE Internet Things J.}, vol. 9, no. 18, pp. 16928--16940, Sept. 2022.
\bibitem{rel9}
M. A. Khan et al., ``A provable and privacy-preserving authentication scheme for UAV-enabled intelligent transportation systems,'' \textit{IEEE Trans. Ind. Informat.}, vol. 18, no. 5, pp. 3416--3425, May 2022.
\bibitem{rel10}
W. Wang, Z. Han, T. R. Gadekallu, S. Raza, J. Tanveer, and C. Su, ``Lightweight blockchain-enhanced mutual authentication protocol for UAVs,'' \textit{IEEE Internet Things J.}, vol. 11, no. 6, pp. 9547--9557, Mar. 2024.
\bibitem{rel11}
T. Alladi, Naren, G. Bansal, V. Chamola, and M. Guizani, ``SecAuthUAV: A novel authentication scheme for UAV-ground station and UAV-UAV communication,'' \textit{IEEE Trans. Veh. Technol.}, vol. 69, no. 12, pp. 15068--15077, Dec. 2020.
\bibitem{rel12}
G. Bansal and B. Sikdar, ``S-MAPS: Scalable mutual authentication protocol for dynamic UAV swarms,'' \textit{IEEE Trans. Veh. Technol.}, vol. 70, no. 11, pp. 12088--12100, Nov. 2021.
\bibitem{rel13}
C. Pu, A. Wall, K. -K. R. Choo, I. Ahmed, and S. Lim, ``A lightweight and privacy-preserving mutual authentication and key agreement protocol for internet of drones environment,'' \textit{IEEE Internet Things J.}, vol. 9, no. 12, pp. 9918--9933, June 2022.
\bibitem{rel14}
R. Karmakar, G. Kaddoum, and O. Akhrif, ``A PUF and fuzzy extractor-based UAV-ground station and UAV-UAV authentication mechanism With intelligent adaptation of secure sessions,'' \textit{IEEE Trans Mob. Comput.}, vol. 23, no. 5, pp. 3858--3875, May 2024.
\bibitem{rel15}
B. Chen et al., ``A security awareness and protection system for 5G smart healthcare based on zero-trust architecture,'' \textit{IEEE Internet Things J.}, vol. 8, no. 13, pp. 10248--10263, July 2021.
\bibitem{rel16}
Y. Bello, A. R. Hussein, M. Ulema, and J. Koilpillai, ``On sustained zero trust conceptualization security for mobile core networks in 5G and beyond,'' \textit{IEEE Trans. Netw. Service Manag.}, vol. 19, no. 2, pp. 1876--1889, June 2022.
\bibitem{rel17}
Y. Liu et al., ``A blockchain-based decentralized, fair and authenticated information sharing scheme in zero trust internet-of-things,'' \textit{IEEE Trans. Comput.}, vol. 72, no. 2, pp. 501--512, Feb. 2023.
\bibitem{rel18}
S. O. Ajakwe, D. -S. Kim, and J. -M. Lee, ``Drone transportation system: Systematic review of security dynamics for smart mobility,'' \textit{IEEE Internet Things J.}, vol. 10, no. 16, pp. 14462--14482, Aug. 2023.
\bibitem{rel19}
Y. Ge and Q. Zhu, ``GAZETA: Game-theoretic zero-trust authentication for defense against lateral movement in 5G IoT networks,'' \textit{IEEE Trans. Inf. Forensics Security}, vol. 19, pp. 540--554, 2024.
\bibitem{rel20}
W. Lei, Z. Pang, H. Wen, W. Hou, and W. Li, ``Physical layer enhanced zero-trust security for wireless industrial internet of things,'' \textit{IEEE Trans. Ind. Informat.}, vol. 20, no. 3, pp. 4327--4336, Mar. 2024.
\bibitem{rel21}
R. Cheng, S. Chen, and B. Han, ``Towards zero-trust security for the metaverse,'' \textit{IEEE Commun. Mag.}, vol. 62, no. 2, pp. 156--162, Feb. 2024.
\bibitem{adpre1}
S. S. M. Chow, S. M. Yiu, L. C. K. Hui, and K. P. Chow, `Efficient forward and provably secure ID-based signcryption scheme with public verifiability and public ciphertext authenticity,'' in \textit{Information Security and Cryptology} (Lecture Notes in Computer Science), vol. 2971. New York, NY, USA: Springer-Verlag, 2004, pp. 352–369.
\bibitem{pre1}
C. Böhm and M. Hofer, ``Introduction,'' in \textit{Physical unclonable functions in theory and practice}, New York, NY, USA: Springer, 2012, ch. 1, pp. 3--37.
\bibitem{pre2}
A. Moubayed, A. Refaey, and A. Shami, ``Software-defined perimeter (SDP): State of the art secure solution for modern networks,'' \textit{IEEE Netw.}, vol. 33, no. 5, pp. 226--233, 2019.
\bibitem{pre3}
S. Nakamoto. \textit{Bitcoin: A peer-to-peer electronic cash system} (2008). [Online]. Available: https://bitcoin.org/bitcoin.pdf

\bibitem{thr1}
D. Dolev and A. Yao, ``On the security of public key protocols,'' \textit{IEEE Trans. Inf. Theory}, vol. 29, no. 2, pp. 198--208, Mar. 1983.

\bibitem{ana1}
 D. Pointcheval and J. Stern, ``Security arguments for digital signatures and blind signatures,'' \textit{J. Cryptol.}, vol. 13, no. 3, pp. 361--396, Mar. 2000.

\bibitem{comp1}
E. Wisse, P. Tedeschi, S. Sciancalepore; and R. Di Pietro, ``A2RID—anonymous direct authentication and remote identification of commercial drones,'' \textit{IEEE Internet Things J.}, vol. 10, no. 12, pp. 10587--10604, June 2023.

\bibitem{comp2}
C. Feng, B. Liu, Z. Guo, K. Yu, Z. Qin, and K. -K. R. Choo, ``Blockchain-based cross-domain authentication for intelligent 5G-enabled internet of drones,'' \textit{IEEE Internet Things J.}, vol. 9, no. 8, pp. 6224-6238, Apr. 2022.

\bibitem{comp3}
S.A. Chaudhry, K. Yahya, M. Karuppiah, R. Kharel, A.K. Bashir, and Y.B. Zikria, ``Gcacs-iod: A certificate based generic access control scheme for internet of drones, '' \textit{Comput. Netw.}, vol. 191, pp. 107999, 2021.

\bibitem{comp4}
M. A. Khan et al., ``A provable and privacy-preserving authentication scheme for UAV-enabled intelligent transportation systems,'' \textit{IEEE Trans. Ind. Informat.}, vol. 18, no.5, pp. 3416--3425, May 2022.


\end{thebibliography}
\end{document}